\documentclass[prl,superscriptaddress,showpacs]{revtex4}
\usepackage{graphicx}
\usepackage{epsf}

\def\bea{\begin{eqnarray}}
\def\eea{\end{eqnarray}}
\def\beq{\begin{equation}}
\def\eeq{\end{equation}}
\newcommand{\ice}[1]{\relax}

\newcommand\VV{\setbox0=\hbox{V}\hbox{\rm V\raise\ht0
  \hbox to0pt{\hss\vbox to0pt{\hbox{v}\vss}}}}
\def\slashchar#1{\setbox0=\hbox{$#1$}           
   \dimen0=\wd0                                 
   \setbox1=\hbox{/} \dimen1=\wd1               
   \ifdim\dimen0>\dimen1                        
      \rlap{\hbox to \dimen0{\hfil/\hfil}}      
      #1                                        
   \else                                        
      \rlap{\hbox to \dimen1{\hfil$#1$\hfil}}   
      /                                         
   \fi}                                         %

\begin{document}

\title{Comment on the new $D_s^{(*)+} \pi^0$ resonances}

\author{Thomas E.~Browder}

\affiliation{Department of Physics and Astronomy, 
University of Hawaii at Manoa, Honolulu, Hawaii 96822}

\author{Sandip Pakvasa}

\affiliation{Department of Physics and Astronomy, 
University of Hawaii at Manoa, Honolulu, Hawaii 96822}

\author{Alexey A.~Petrov}

\affiliation{Department of Physics and Astronomy, 
Wayne State University, Detroit, MI 48201}

\affiliation{Michigan Center for Theoretical Physics,
University of Michigan, Ann Arbor, MI 48109}

\begin{abstract}
We propose an explanation of the new resonances
observed in $D_s^{(*)+} \pi^0$ decays. We suggest that the
data can be explained by the mixing of conventional
p-wave excited $D_s^+$ mesons with 4-quark states.
The narrow states observed in $D_s^+ \pi^0$ and
$D_s^{*+}\pi^0$ are primarily p-wave $D_{sJ}^{*}$ states,
while the predominantly 4-quark states are shifted above $D^{(*)} K$ threshold
and should be broad.
Ranges for the mixing parameter and mass of the 4-quark
state in this scenario are given. Other experimental consequences
are discussed.
\end{abstract}


\maketitle



Meson spectroscopy is an important laboratory for understanding quark 
confinement. Mesons containing one heavy quark can provide 
invaluable information about the structure of the QCD Lagrangian,
as spectroscopic considerations simplify significantly in the 
limit of infinitely heavy quark. In this limit the heavy quark spin $S_Q$ 
decouples, so the total angular momentum of the light degrees of freedom 
$J_l^p$ becomes a ``good'' quantum number. This leads to an important
prediction of heavy quark symmetry: the appearance of heavy meson 
states in the form of degenerate parity doublets classified by the 
total angular momentum of the light degrees of freedom. This mass degeneracy 
is lifted with the inclusion of subleading $1/m_Q$ corrections. This useful 
picture is built into many quark-model descriptions of heavy meson spectra. 
The resulting models have been very successful in explaining the spectrum 
of negative-parity scalar and vector $J_l^p=1/2^-$ and positive-parity 
vector and tensor $J_l^p=3/2^+$ states.

A narrow resonance in $D_s^+\pi^0$ was recently reported by BaBar~\cite{dspi0_babar} 
and confirmed by the 
CLEO~\cite{dspi0_cleo} and Belle~\cite{dspi0_belle} collaborations. Its decay 
patterns suggest a quark-model $0^+$ classification, which would 
identify it with the positive-parity $J_l^p=1/2^+$ $p$-wave state. As in the 
$D$ meson system, $p$-wave states for the $D_s^+$ system are expected, and two 
narrow states, $D_{s1}(2536)$ and $D_{s2}(2573)$ were discovered by ARGUS and 
CLEO, respectively~\cite{PDG}. 
In analogy to the $D$ system, two broad states are also expected.

The mass of the new state $2317.6\pm 1.3$ MeV appears surprisingly low for quark model 
practitioners. In fact, this state appears below $DK$ threshold, closing off the most 
natural decay channel for this state. This forces it to decay mainly via an
isospin-violating transition into the $D_s^+\pi^0$ final state and makes its width quite 
narrow. Its mass disagrees with most existing predictions of quark 
models~\cite{godfrey_isgur,eichten,Gupta:1994mw,Zeng:1994vj,Kalashnikova:2001ig,Ebert:1997nk}.
For example, a mass of $2487$ MeV is obtained in the recent potential model 
calculation by Eichten and Di Pierro~\cite{eichten}. Quenched lattice calculations 
also seem to favor larger values of the mass of this state~\cite{Lewis:2000sv} 
(see, however, \cite{lattice}). 
This led to a lively discussion of the possible non-$q\bar q$ nature
of this state~\cite{barnes_lipkin,cheng_hou,szcz,cahn_jackson}.
In addition, a second narrow state is observed in $D_s^{*+}\pi^0$ at a mass near
2460 MeV~\cite{dspi0_cleo,dspi0_belle}. 
This state would naturally be identified as a spin 1 positive parity 
$p$-wave meson. However, its mass also appears too low for the potential model 
expectations (e.g. 2605 MeV~\cite{eichten}).

The observed low values of the masses for these states, however, do
not signal a breakdown of quark-model descriptions of the heavy meson 
spectrum, as it is difficult to assess the accuracy of these
predictions, especially in the charm sector. Many authors make use of the 
non-relativistic nature of the charm quark, taking into account
$1/m_c$ corrections only. For the $0^+$ state, quark model predictions 
range from the values of $2387-2395$~MeV~\cite{Gupta:1994mw,Kalashnikova:2001ig} 
(still above the $DK$ threshold) on the low end of the spectrum to 
$2508$~MeV~\cite{Ebert:1997nk} on the high end. Since the described 
phenomena are highly non-perturbative, one should be careful before 
making a judgment on the nature of a given state based solely on the 
prediction of a given quark model. For example, as discussed above, in 
the $m_c \to \infty$ limit the $0^+$ and $1^+$ states are expected to 
become degenerate in mass, $m_{0^+},m_{1^+} \to M$. This can be
emulated in quark models by neglecting heavy-quark symmetry-violating $1/m_c$ 
corrections. Yet, different quark models predict very different
behavior in this "heavy-quark limit": for instance, one potential 
model~\cite{Gupta:1994mw} predicts that the mass $M$ of the 
$(0^+, 1^+)$ multiplet will decrease to approximately $2382$~MeV
(which is less than the mass of the $0^+$ state predicted in this model with
the full potential), while in a QCD string
model~\cite{Kalashnikova:2001ig} 
it is expected to increase up to $2500$~MeV (which is much greater
than the mass of the $0^+$ state predicted in this model with the full 
potential). In addition, quark models, modified to include 
chiral symmetry constraints, generally predicted lower values of mass 
splitting between $(0^-,1^-)$ and $(0^+,1^+)$ multiplets, of the order
of $200-300$~MeV~\cite{chiral}. Nevertheless, the fact that the new state 
appears below $D K$ threshold and is almost degenerate with a
non-strange $0^+$ $p$-wave $D$ state~\cite{ddst_belle} is curious and deserves an 
investigation.

\begin{figure}[htb]
\centerline{
\epsfxsize 3.0 truein \epsfbox{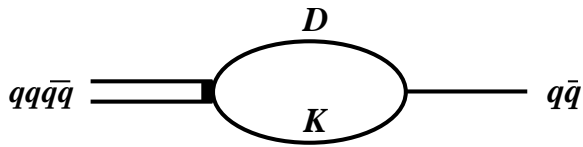}
 }
\caption{Example of the interaction generating mixing of the $q\bar{q}$ and
4-quark states. In general, all possible (multibody) intermediate
states such as $D^{(*)}K^{(*)}, DK\pi\pi,...$ will contribute.}
 \label{fig1}
 \end{figure}

It has been shown that the proper description of the low lying scalar mesons
requires the inclusion of the coupling of the scalar $q \bar{q}$ states to
the s-wave meson-meson channel, and this coupling is very
important~\cite{beveren}. By an extension of this argument, we would expect 
the scalar $0^+$ $c\bar{s}$ state to mix with the $DK$ $s$-wave state near threshold.
This mixing can occur quite naturally if the states have the same quantum numbers
(for instance, via the common intermediate states, see Fig.~\ref{fig1}) and 
could shift the mass of the $c\bar{s}$ state.

Alternatively, the possibility of 4-quark states with quark content $c \bar{s} q
\bar{q}$ has been  discussed in the past~\cite{lipkin_1,suzuki}. 
Here we consider the possibility that there is a 4-quark state that mixes 
with the $c \bar{s}$ state and shifts its mass below $D K$ threshold.

We briefly review two-state mixing for the case of two nearby
hadronic states. Let $m_0$ be the unshifted mass of the $q\bar q$
state and $\tilde{m_0}$ be the unshifted mass of the 4-quark state. 
The strong interaction can mix these states by introducing non-diagonal 
matrix elements $\delta$ in the mass matrix of $qq\overline{qq}$ and $q\bar{q}$
states, which has to be diagonalized. The 
masses of the two mixed (mass eigen)states, $m_1$ and $m_2$ will then be 
given by 
\begin{eqnarray}
m_1 = {1 \over 2} \Sigma + {1 \over 2} \sqrt{\Delta^2 + 4\delta^2} \\
m_2 = {1 \over 2} \Sigma - {1 \over 2} \sqrt{\Delta^2 + 4\delta^2} 
\end{eqnarray}
where $\Sigma = m_0 + \tilde{m_0}$ and $\Delta = m_0 - \tilde{m_0}$.
The mixing angle that defines the composition of resulting mass eigenstates 
in terms of the original $qq\overline{qq}$ and $q\bar{q}$ states is given by 
$\tan (2\theta) = 2 \delta/\Delta$.

\begin{figure}[htb]
\centerline{
\epsfxsize 4.0 truein \epsfbox{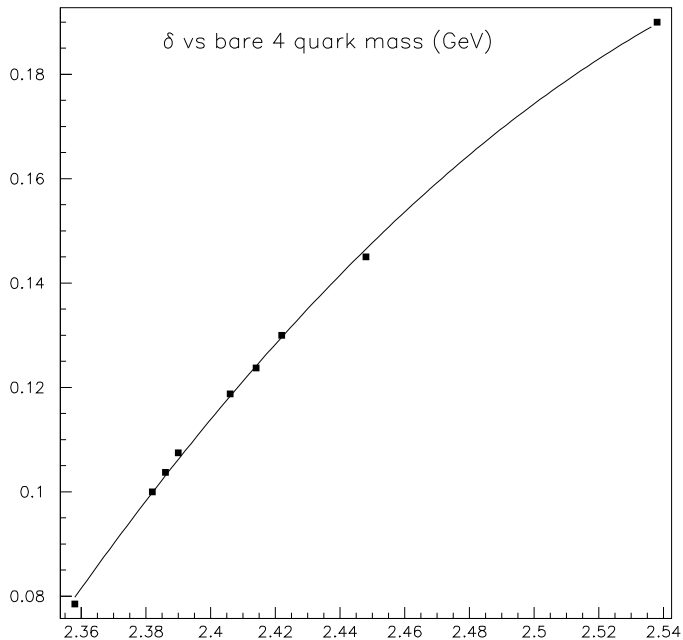}
 }   
\caption{$\delta$  versus $\tilde{m_0}$ 
for solutions that satisfy the constraints of this scenario.}        
 \label{mix_figure}
 \end{figure}

Let us be more specific. For example, take
$\tilde{m_0}$ the bare mass of the 4-quark state to be just above
$D K$ threshold at 2.37 GeV and $m_0$  at the value 2.48 GeV as given
by the potential model; then for a mixing parameter $\delta$
of  0.092 GeV, we find two mixed states with masses
$m_2=2.3194$ GeV and $m_1=2.5375$ GeV. The corresponding mixing angle 
is $28.8^{0}$. The lower state, which is dominantly a p-wave $D_s$ meson,
 is below $D K$ threshold, 2.358 GeV, but above $D \pi$ threshold, 2.103 GeV.
Conversely, the higher state, which is dominantly a 4-quark
system, is above $D K$ threshold and is broad.
There is a large set of parameters for the bare 4 quark mass
and mixing that can shift the mass of the $D_{sJ}^{*}$ to the observed
value of $2317$~GeV. This is illustrated in Figure~\ref{mix_figure}. 
If $\tilde{m_0}$ is too high, it is difficult to shift the 
$D_{sJ}^{*}$ state, with a reasonable amount of mixing.

In principle, we also expect the same type of mixing to occur for the $1^+$ state.
For instance, a molecular-type bound-state of
$D^*K$ of the kind suggested by Lipkin~\cite{lipkin_1,barnes_lipkin} 
and others, can mix with the $c\bar{s}$ $^3P_1$ state and shift 
the $1^+$ states as well. To calculate
the allowed range of mixing in this case is more complicated since there
is mixing between the two $1^+$ states in addition to the mixing
between the lower $1^+$ state and the relevant 4 quark state, and we do not 
attempt a detailed fit.


The properties and decay modes of the lower mixed states should be very similar
to those expected for $p$-wave $D_s$ mesons. 
Like the broad $D^{**}$ states perhaps some of the additional states may be identified
in exclusive $B$ decays such as $B\to D^{(*)0}\bar{D}^0 K$
with enough statistics.
The observation of these extra states could lend extra support for the existence 
of the multiquark and/or molecular states. 

The 4 quark states can also give rise to very distinctive final states
that are doubly charged~\cite{lipkin_1,suzuki}. For example, there may be
$I=1$ states including a doubly charged state and a neutral state
with decay modes into $D^+ K^+$, etc. In the scenario described
here, these exotic final states are above $D K$ threshold and are
therefore quite broad, experimentally difficult to find or even may
be non-existent as "real" bound states. Note that we do not expect narrow doubly
charged $D_s^+ \pi^+$ or $D_s^+\pi^-$ states in the scenarios
described above.

It has also been suggested~\cite{barnes_lipkin,Godfrey:2003kg} that
radiative transitions of the type $D^*_{sJ}(2317) \to D_s^* \gamma$
could be used to test the nature of these states. In particular, it was 
suggested that a very small branching ratio for the radiative
decays would be an indication of a non-$q\bar{q}$ nature of these
states. In the scheme described above this test will be useful
only for very small or very large (about $\pi/2$) values of the mixing 
angle between $c\bar{s}$ and $c\bar{s}q\bar{q}$ states, as the 
contribution of $c\bar{s}q\bar{q} \to D_s^* \gamma$ transition is 
suppressed by $\sin\theta$. This is not
so for moderate values of the mixing angle~\cite{OurRadProject}.


In conclusion, we propose that the new states observed in
$D_s^+\pi^0$ and $D_s^{*+}\pi^0$ are shifted by mixing with 4-quark
states. We expect that the observed states will have properties similar
to those expected for $p$-wave $D_s$ mesons except for their masses. 
The additional states responsible for the mixing and mass shifts
will be above $D^{(*)} K$ threshold and have large widths.

We thank M.~Suzuki for useful discussions. 
We acknowledge support from the U.S.\ Department of Energy under
grants DE-FG03-94ER40833, DE-FG02-96ER41005.


\end{document}